\documentclass[natbib,twocolumn,twoside]{svmultiag}
\usepackage{graphicx}
\usepackage{multirow}
\graphicspath{{./figures/}}
\title*{Observations of radio sources near the Sun}
\titlerunning{Observations of radio sources near the Sun}

\author{O.~Titov, S.~Lambert, B.~Soja, F.~Shu, A.~Melnikov, J.~McCallum, L.~McCallum, M.~Schartner, A.~de Witt, D.~Ivanov, A.~Mikhailov, S.~O.~Yi, W.~Chen, B.~Xia, M.~Ishigaki, S.~Gulyaev, T.~Natusch, S.~Weston }
\authorrunning{O.~Titov et al.}

\institute{Oleg Titov \at
Geoscience Australia,
PO Box 378
Canberra
2601 Australia
\and
S\'ebastien Lambert \at
SYRTE, Observatoire de Paris, Universit\'e PSL, CNRS, Sorbonne Universit\'e, LNE, Paris, France
\and
Benedikt Soja \at
Jet Propulsion Laboratory, California Institute of Technology, 4800 Oak Grove Drive, Pasadena, CA 91109, USA
\and
Fengchun Shu, Wen Chen abd Bo Xia \at
Shanghai Astronomical Observatory, 80 Nandan Road, Shanghai, 200030, China
\and
Alexei Melnikov, Dmitrii Ivanov and Andrei Mikhailov \at
Institute of Applied Astronomy, Kutuzov Embankment, 10, Saint-Petersburg, 191187, Russia
\and 
Jamie McCallum and Lucia McCallum \at
University of Tasmania, Private Bag 37, Hobart, Tasmania, 7001, Australia
\and
Aletha de Witt \at
Hartebeesthoek Radio Astronomy Observatory, PO Box 443, Krugersdorp, 1740, South Africa
\and
Matthias Schartner \at
Department of Geodesy and Geoinformation, Research Group Advanced Geodesy, TU Wien, Gusshausstra{\ss}e 27-29/E120.4, Wien-1040, Austria
\and
Sang Oh Yi \at
National Geographic Information Institute, Space Geodetic Observatory, Sejong, PO Box 30060, South Korea
\and
Masafumi  Ishigaki \at
Geospatial Information Authority of Japan 1, 
Kitasato, Tsukuba 
305-0811, Japan
\and
Sergei Gulyaev, Tim Natusch and Stuart Weston \at
Institute for Radio Astronomy and Space Research,
Auckand University of Technology,
Auckland,
1010, New Zealand
}
\begin{document}
\maketitle
%----------------------------------------------------------------------------
\abstract{
Geodetic Very Long Baseline Interferometry (VLBI) data are capable of measuring the
light deflection caused by the gravitational field of the Sun and large planets
with high accuracy. The parameter $\gamma$ of the parametrized Post-Newtonian (PPN) 
formalism estimated using observations of reference radio sources near the Sun should be 
equal to unity in the general relativity. 
We have run several VLBI experiments tracking reference radio sources from 1 to 3 degrees from the Sun.
The best formal accuracy of the parameter $\gamma$ achieved in the single-session mode 
is less than 0.01 percent, or better than the formal accuracy obtained with a global solution 
included all available observations at arbitrary elongation from the Sun.  
We are planning more experiments starting from 2020 using better observing conditions near the minimum of 
the Solar activity cycle.
}


%----------------------------------------------------------------------------
\section{Introduction}                                
\label{SEC:introduction}
%----------------------------------------------------------------------------

%Very long baseline interferometry (VLBI) measures the difference in arrival times (known as group delay) of radio waves at two radio telescopes from distant radio sources with a precision of 20-40 ps \citep{SchuhBehrend2012}. The observations are carried out at two frequency bands: 2.3~GHz (S-band) and 8.4~GHz (X-band); the lower frequency is used to calibrate ionospheric fluctuations in X-band data. By combining many years of observations, this technique is capable of producing very accurate positions of the reference radio sources. E.g., the error floor of the current realization of the fundamental celestial reference frame, the ICRF3, is 30~$\mu$as \citep{CharlotJacobsGordon2019}. For one standard single geodetic VLBI experiment, positions of radio sources are estimated with an accuracy of about 0.1 to 1~mas.

In accordance with General Relativity the radio waves slow down due to the gravitational potential of the Sun \citep[the so-called Shapiro effect; see][]{Shapiro1964,Shapiro1967}, making very long baseline interferometry (VLBI) a useful tool for testing General Relativity by means of the parameterized post-Newtonian (PPN) formalism \citep{Will1993}. Nevertheless, the accuracy of the PPN parameter $\gamma$ obtained from absolute or differential VLBI observations \citep{FomalontKopeikinLanyi2009,LambertLePoncin-Lafitte2009,LambertLePoncin-Lafitte2011} remains worse than the current best limit of $(2.1\pm2.3)\times10^{-5}$ based on Cassini radio science experiment \citep{BertottiIessTortora2003} by an order of magnitude. The upper limits on the parameter $\gamma$ have been improved substantially in the past 30 years \citep{RobertsonCarter1984,RobertsonCarterDillinger1991,LebachCoreyShapiro1995,FomalontKopeikin2003}, but some authors \citep{ShapiroDavisLebach2004,LambertLePoncin-Lafitte2009} found degradation in the estimates of $\gamma$ with elongation, and suggested that this systematic effect may limit the improvement in the VLBI-derived $\gamma$ upper limits, despite the dramatic growth in the number of observations in recent decades.

The current paper focuses on radio source approaches at angular distances less than three degrees from the centre of the Sun in order to measure the light deflection effect at the highest magnitude and, thus, to avoid a possible bias caused by observations at larger elongations. We report on two special VLBI sessions, AUA020 (May 2017) and AOV022 (May 2018), on the single-session estimates of $\gamma$.

%----------------------------------------------------------------------------
\section{Data}                                 
\label{SEC:Data}
%----------------------------------------------------------------------------

\begin{figure}[t!]
\includegraphics[width=0.5\textwidth]{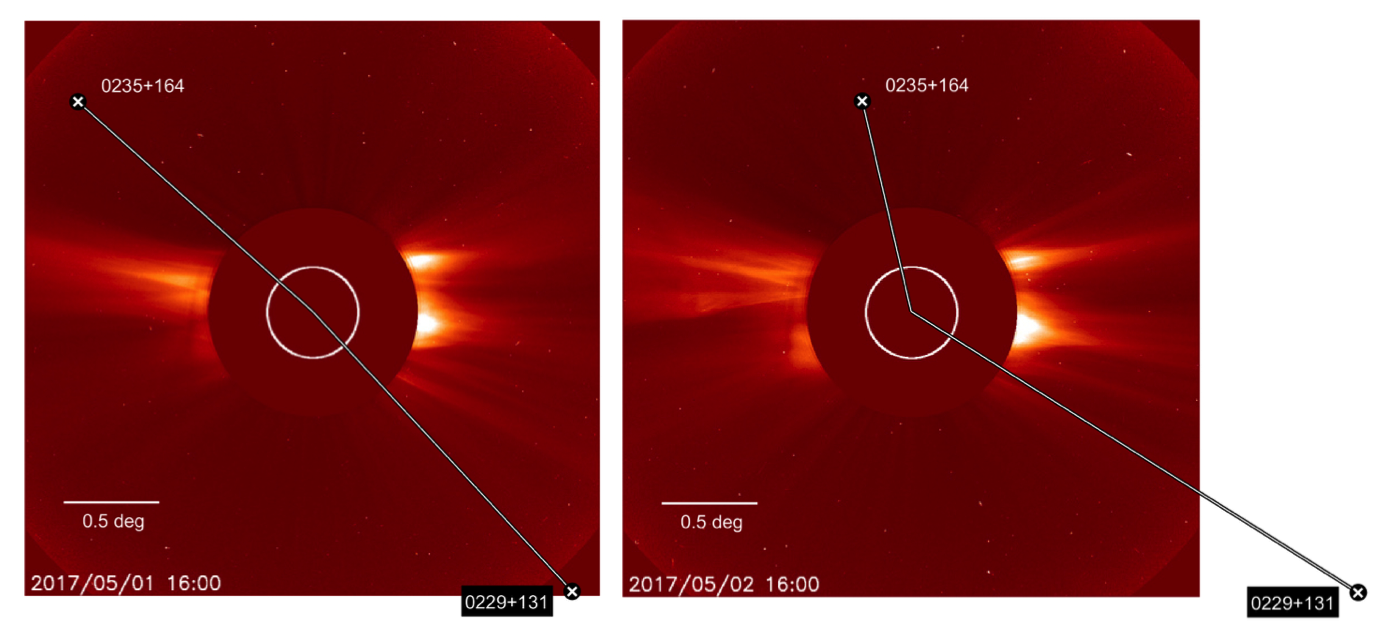}
\caption{
Geometry of the radio sources close to the sun at the start (Left) and at the end (Right) of VLBI session AUA020 with respect to a LASCO C2 image of the solar corona. The Sun is hidden behind the occultation disc of the coronagraph, with the white inner circle representing the limb of the Sun. The field-of-view is 1.5 degrees elongation.}
\label{Fig_Sun}
\end{figure}

A dedicated geodetic VLBI experiment (AUA020, 01-02 May, 2017, part of AUSTRAL program) was scheduled to probe the gravitational delay effect using a network of seven radio telescopes (Svetloe, Zelenchukskaya, Badary, HartRAO, Seshan25, Sejong, and Hobart26). Two radio sources 0229+131 and 0235+164 were observed at range of angular distances from 1.15$^\circ$ to 2.6$^\circ$ from the Sun. The position of both radio sources with respect to the Sun at the start of the experiment is shown on Fig.~\ref{Fig_Sun}. A serious issue in such a configuration is the solar thermal noise that penetrates to the signal through the side lobes, and could cause loss of data due to striking the signal-to-noise ratio. To overcome the problem, one has to

1) select strong radio sources with larger correlated flux density in both frequency bands,

2) use large radio telescopes with narrow side lobes and better sensitivity, and

3) use the highest possible data rate recording (e.g., 1 Gbps) to gain a better signal-to-noise ratio during the same integration time.

More details about the schedule design are publsihed in \citep{Titov2018}.

%............................................................................
% FIG:02 START
%............................................................................
%............................................................................
% FIG:02 END
%............................................................................

The target radio source 0229+131 is a defining source of the ICRF2 whose position is given with an accuracy close to the ICRF2 noise floor of 40~$\mu$as. The position of the second target 0235+164 is less accurate by a factor of five but still at the level of the ICRF2 median error and largely below the millisecond of arc. Both sources are compact and their structure indices measured at the time of the ICRF2 work were of 2.4 and 1.3, respectively, ensuring a structure delay lower than 2~ps \citep{FeyCharlot1997}.

%............................................................................
% TAB:01 START
%............................................................................
%\begin{table}
%\centering
%\caption{blabla blublu bloblo}
%\begin{tabular}{lll}
%\hline
%blabla & blublu & bloblo \\
%\hline
%1 & 2 & 3 \\
%3 & 2 & 1 \\
%2 & 3 & 1 \\
%\hline
%\end{tabular}
%\label{TAB: blabla}
%\end{table}
%............................................................................
% TAB:01 END
%............................................................................

%---------------------------------------------------------------------------
\section{Analysis and Results}                                         
\label{SEC:analysis and results}
%---------------------------------------------------------------------------
For purpose of cross-checking the results and testing their robustness, we processed the VLBI session AUA020 within two independent teams with two independent geodetic VLBI analysis software packages. The duplication of the analyses with two software packages also allows to use some specific options that are available on only one of them. The first analysis package is OCCAM \citep{2004ivsg.conf..267T} that implements the least-squares collocation method \citep{2000ITN_Titov} for calibrating the wet troposphere fluctuations, and to account for the mutual correlations between observables. The second one is Calc/Solve \citep{MaClarkRyan1986}, developed and maintained by the geodetic VLBI group at NASA GSFC, that uses classical least-squares. More details about the data analysis design are discussed in \citep{Titov2018}

Table 1 shows the results of the AUA020 experiment data analysis. 
Uncertainties on $\gamma$ lie between $0.9\times10^{-4}$ and $4\times10^{-4}$. Our estimates appear therefore as precise as that obtained from global solutions using thousands of VLBI experiments \citep{LambertLePoncin-Lafitte2009,LambertLePoncin-Lafitte2011}. The formal error is about two times lower when $\gamma$ is fitted to the observations of the radio source that is two times closer to the centre of the Sun (0235+164) than to that of its counterpart (0229+131). Using all scans returns a result similar to using only scans relevant to 0229+131 and 0235+164, confirm that only sources a low elongation can efficiently constrain the PPN parameter. Solutions from both software packages are consistent within the standard errors. The difference of postfit rms between OCCAM and Calc/Solve might find its origin in the different modeling of the nuisance parameters (stochastic versus CPWL function).  No large systematics are detected except a $2.7\sigma$ deviation in the Calc/Solve solution when only 0229+131 is used and whose origin is unclear: as both solutions started from the same a priori, the issue could rather be in the estimation method or in the handling of troposphere/clock parameters.

It appears that during session AUA020, data in three channels at Sejong station were lost due to technical reasons. Therefore, we reprocessed the previous analyses after downweighting (but not suppressing) Sejong data. (We could test this option with OCCAM only since Calc/Solve does not handle downweighting.) The postfit rms of the solution is significantly lowered. The formal error on $\gamma$ is marginally lowered down to $9\times10^{-5}$.

\begin{figure}[t!]
\includegraphics[width=0.5\textwidth]{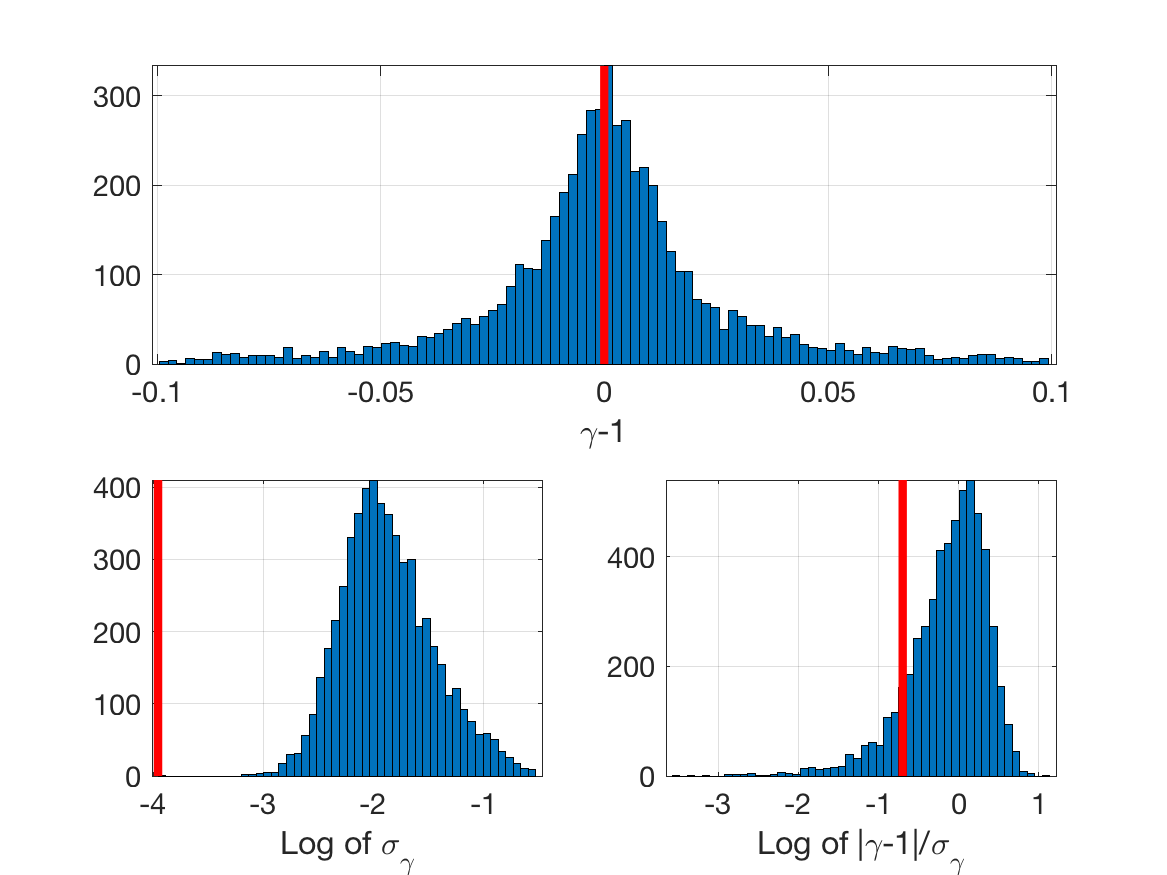}
\caption{
Distributions of (Top) estimates of $\gamma-1$, (Bottom-left) their formal errors, and (Bottom-right) normalized estimates of $\gamma-1$ for all of the geodetic VLBI sessions. The vertical, red bar stands for the results of the AUA020 session.}
\label{figfulldb}
\end{figure}

For purpose of comparison of the AUA020 session with other standard geodetic VLBI sessions, we estimated $\gamma$ with Calc/Solve using the parameterization described above for each of sessions of the full geodetic VLBI data base made available by the International VLBI Service for geodesy and astrometry (IVS) since 1979 (at the exclusion of intensive sessions). The median postfit rms is 27 ps that is close to the postfit rms of the AUA020 session. The distribution of the obtained values of $\gamma-1$ is 
shown in Fig.~\ref{figfulldb} along with distributions of errors and normalized estimates. The distribution of errors in log-scale is slightly asymmetric, exhibiting a `tail' on its right side that might traduce results from sessions not designed for precise astrometry. Nevertheless, assuming a Gaussian shape, the log-scaled distribution peaks at 10$^{-2}$ with a $\sigma$ of $\sim0.5$. This makes the error estimate from AUA020, that is two orders of magnitude less, somewhat `outstanding'. The bottom-right panel of Fig.~\ref{figfulldb} shows that the major part of the sessions does not bring severe systematics, the estimates of $\gamma$ being unity within the error bars; session AUA020 is part of the session group that presents the lowest systematics.

%We also processed solutions parameterized as in \cite{LambertLePoncin-Lafitte2009,LambertLePoncin-Lafitte2011}, thus estimating $\gamma$ as a global parameter over the same 6301 sessions, totaling 12.6 millions of ionosphere-free group delays. A priori positions for radio sources were taken from the ICRF2 \citep{FeyGordonJacobs2015} and a no-net rotation constraint was applied to the defining sources. The postfit rms delay of the solution is 26 ps and its $\chi^2$ per degree of freedom is 1.00. We obtained $\gamma-1=(2.72\pm0.92)\times10^{-4}$ that yields a slight improvement with respect to \cite{LambertLePoncin-Lafitte2011} mainly due of the $\sim$5.3 millions of observations accumulated in the mean time. Removal of the session AUA020 led to $\gamma-1=(2.57\pm0.97)\times10^{-4}$, showing that AUA020 marginally - but still at a detectable level - improves the formal error at the level of $5\times10^{-5}$. However, these global solutions exhibit systematics at the level of 2-3$\sigma$ that may have their origin in spurious or unmodeled deformations of the celestial reference frame.

\begin{table}
\caption{Estimates of $\gamma-1$ for the session AUA020, in unit of 10$^{-4}$, along with the session $\chi^2$ and the postfit rms delay $r$ in ps.}
\label{tabsol}
\centering
\begin{tabular}{lrrrrr}        
\hline
\hline
\noalign{\smallskip}
& & $\gamma-1$ & $\sigma_{\gamma}$& $\chi^2$ & $r$ \\
& & $10^{-4}$ & $10^{-4}$& & \\
\noalign{\smallskip}
\hline
\noalign{\smallskip}
\multirow{10}{*}{OCCAM} & \multicolumn{5}{l}{All stations} \\ \cline{2-6}
\noalign{\smallskip}
& All scans        &   0.56 & 1.15 & 0.34 & 28 \\
& 0235+164          &   1.34 & 1.58 & 0.34 & 28 \\
& 0229+131          &  -1.54 & 3.41 & 0.34 & 28 \\
& Both                   &   0.53 & 1.14 & 0.34 & 28  \\
\noalign{\smallskip}
& \multicolumn{5}{l}{With Sejong downweighted} \\ \cline{2-6}
\noalign{\smallskip}
& All scans           &   0.91 & 0.94  & 0.27 & 21 \\
& 0235+164            &   1.64 & 1.29  & 0.27 & 21 \\
& 0229+131            &   0.32 & 2.83  & 0.27 & 21 \\      
& Both                     &   0.89 & 0.94  & 0.27 & 21 \\
\noalign{\smallskip}
\hline
\noalign{\smallskip}
\multirow{4}{*}{Calc/Solve}
& All sources                         &  -0.22 & 1.10 & 0.84 & 26 \\
& 0235+164                          &   1.85 & 1.48 & 0.84 & 26 \\
& 0229+131                          &  -6.84 & 2.53 & 0.84 & 26 \\      
& Both                                   &  -0.26 & 1.09 & 0.84 & 26 \\
\noalign{\smallskip}
\hline
\end{tabular}
\end{table}

%............................................................................
% FIG:03 START
%............................................................................
%............................................................................
% FIG:03 END
%............................................................................

\begin{figure}[t!]
\includegraphics[width=0.5\textwidth]{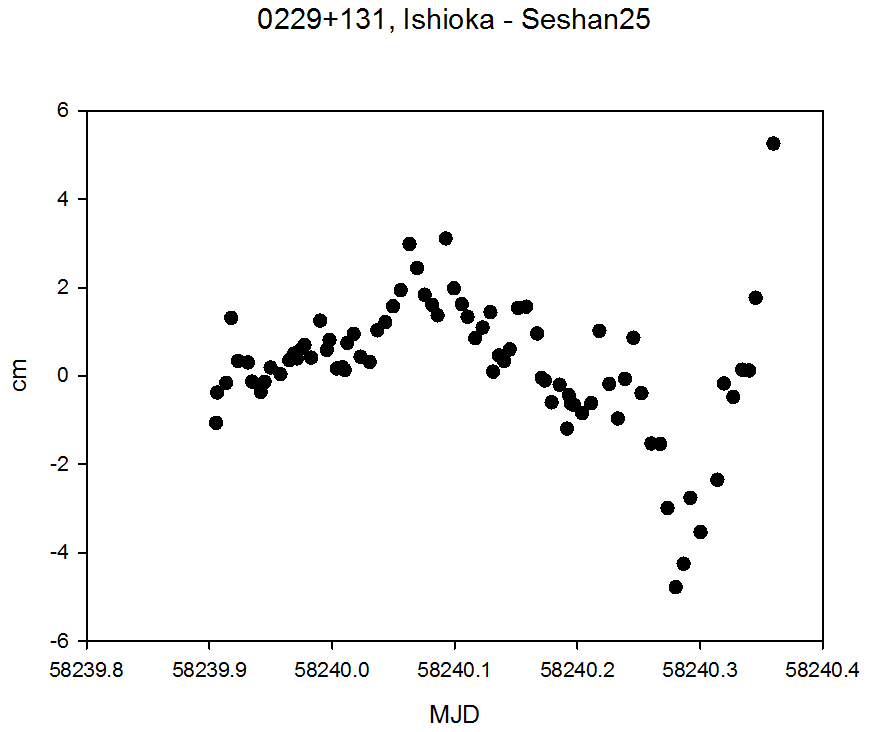}
\includegraphics[width=0.5\textwidth]{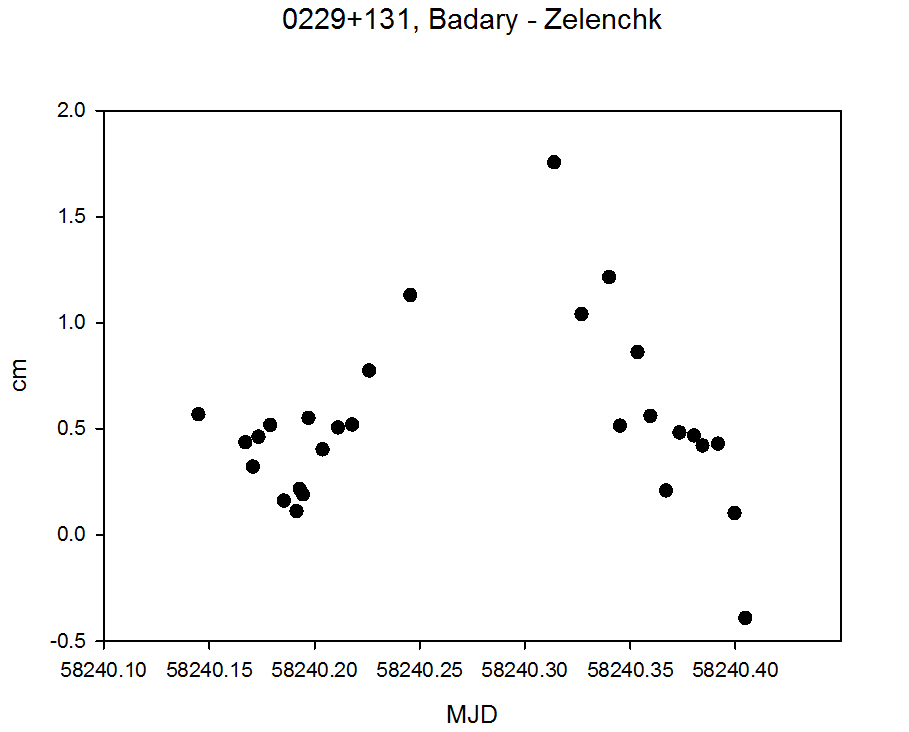}
\caption{Post-fit residuals of 0229+131 with baselines Ishioka-Seshan25 and Badary-Zelenchk.}
\label{aov022_res}
\end{figure}

Another experiment (AOV022) was undertaken on 01-02 May, 2018 with ten radio telescopes (Svetloe, Zelenchukskaya, Badary, Hobart26, Seshan25, Kunming, Ishioka, Yarragadee, Katherine, Warkworth). The same radio sources (0229+131 and 0235+164) were scheduled with the same strategy. 
The statistics of the result was found to be 2-3 times worse than from AUA020, presumably, due to severe source structure delay effect. Fig.~\ref{aov022_res} shows the post-fit residuals of for radio source 0229+131 for two baselines, Ishioka-Seshan25 and Badary-Zelenchk. The variations of the residuals are consistent to the variations induced by the source structure 
\citep{TitovLopez2018} therefore, we believe, that more detailed analysis is required to obtain a better statistic for this experiment.

%............................................................................
% FIG:03 START
%............................................................................
%............................................................................
% FIG:02 END
%............................................................................

%---------------------------------------------------------------------------
\section{Conclusions and outlook}                                 
\label{SEC:conclusions}
%---------------------------------------------------------------------------

   In this paper we discuss our results on testing of general relativity with geodetic VLBI using the close approach of the Sun to the
reference radio sources. It was a general misconception in the past that the effect of the plasma of the solar corona completely disturbs the interferometric responce for light rays passing within several degrees from the Sun.
We proved that these perturbations are at an acceptable level unless the signals pass active streamers in the solar corona.
Therefore, the standard dual frequency calibration facilitates the stochastic noise induced by charged particles the solar corona by the same way as for the ionosphere around the Earth. 

While the systematic effects based on radial or dipole models of the corona appear to be negligible, individual group delay observations are affected by random scatter caused by small-scale coronal structure and temporal variations thereof. Since these perturbations do not systematically affect the observations, we assume that they cancel out over the period of observations (17 hours with observations angularly close to the Sun). Since the ray paths to the radio sources 0235+164 and 0229+131 within small solar elongation happened to be in quiet regions (cf. Fig.~\ref{Fig_Sun}), the scatter was small enough that precise group delays could be successfully determined at such small elongations.

The major source of stochastic noise in VLBI measurements resides in the unknown wet troposphere delay. The difference between VLBI estimates of the wet troposphere delay and independent radiometer data appears to stay within 3 mm, or 10 ps \citep{TitovStanford2013} suggesting that the impact of the wet troposphere delay on the astrometric light deflection angle estimation near the Sun is negligible.

Overall, a total improvement of the uncertainty on $\gamma$ by a factor of ten is expected, enabling to challenge the current limit imposed by the Cassini radio science experiment of \cite{BertottiIessTortora2003}, although the Gaia astrometry on Solar system objects is expected to deliver an accuracy of $10^{-6}$ \citep{MignardKlioner2009}.

%---------------------------------------------------------------------------
\section{Acknowledgments}                                 
\label{SEC:acknowledgements}
%---------------------------------------------------------------------------

This paper is published with the permission of the CEO, Geoscience Australia. B. Soja's research was supported by an appointment to the NASA Postdoctoral Program, administered by Universities Space Research Association, at the Jet Propulsion Laboratory, California Institute of Technology, under a contract with National Aeronautics and Space Administration. We are grateful to D. Gordon (GSFC) for post-processing reduction of the AUA020 data.

%\bibliographystyle{spbasic}
%\bibliography{/Users/seb/Documents/Brol/seb}

\begin{thebibliography}{26}
\providecommand{\natexlab}[1]{#1}
\providecommand{\url}[1]{{#1}}
\providecommand{\urlprefix}{URL }
\expandafter\ifx\csname urlstyle\endcsname\relax
  \providecommand{\doi}[1]{DOI~\discretionary{}{}{}#1}\else
  \providecommand{\doi}{DOI~\discretionary{}{}{}\begingroup
  \urlstyle{rm}\Url}\fi
\providecommand{\eprint}[2][]{\url{#2}}

\bibitem[{{Bertotti} et~al(2003){Bertotti}, {Iess}, and
  {Tortora}}]{BertottiIessTortora2003}
{Bertotti} B, {Iess} L, {Tortora} P (2003) {A test of general relativity using
  radio links with the Cassini spacecraft}. Nature 425:374--376
  
\bibitem[{{Charlot} et~al(2019){Charlot}, {Jacobs}, {Gordon}, {Lambert},
  {Boehm}, {de Witt}, {Fey}, {Heinkelmann}, {Skurikhina}, {Titov}, {Arias},
  {Bolotin}, {Bourda}, {Ma}, {Malkin}, {Nothnagel}, {Gaume}, {Mayer}, and
  {MacMillan}}]{CharlotJacobsGordon2019}
{Charlot} P, {Jacobs} CS, {Gordon} D, et al. (2019) {{The third realization of the International Celestial
  Reference Frame by very long baseline interferometry}}. Astronomy \&
  Astrophysics (in preparation)

\bibitem[{{Fey} and {Charlot}(1997)}]{FeyCharlot1997}
{Fey} AL, {Charlot} P (1997) {VLBA Observations of Radio Reference Frame
  Sources. II. Astrometric Suitability Based on Observed Structure}.
  Astrophysical Journal 111:95--142

\bibitem[{{Fey} et~al(2015){Fey}, {Gordon}, {Jacobs}, {Ma}, {Gaume}, {Arias},
  {Bianco}, {Boboltz}, {B{\"o}ckmann}, {Bolotin}, {Charlot}, {Collioud},
  {Engelhardt}, {Gipson}, {Gontier}, {Heinkelmann}, {Kurdubov}, {Lambert},
  {Lytvyn}, {MacMillan}, {Malkin}, {Nothnagel}, {Ojha}, {Skurikhina},
  {Sokolova}, {Souchay}, {Sovers}, {Tesmer}, {Titov}, {Wang}, and
  {Zharov}}]{FeyGordonJacobs2015}
{Fey} AL, {Gordon} D, {Jacobs} CS, {Ma} C, et al. (2015) {The Second Realization of the International
  Celestial Reference Frame by Very Long Baseline Interferometry}. Astronomical
  Journal 150:58

\bibitem[{{Fomalont} et~al(2009){Fomalont}, {Kopeikin}, {Lanyi}, and
  {Benson}}]{FomalontKopeikinLanyi2009}
{Fomalont} E, {Kopeikin} S, {Lanyi} G, {Benson} J (2009) {Progress in
  Measurements of the Gravitational Bending of Radio Waves Using the VLBA}.
  Astrophysical Journal 699:1395--1402
  
\bibitem[{Fomalont and Kopeikin(2003)}]{FomalontKopeikin2003}
Fomalont EB, Kopeikin SM (2003) The measurement of the light deflection from
  jupiter: Experimental results. The Astrophysical Journal 598(1):704
  
\bibitem[{{Lambert} and {Le
  Poncin-Lafitte}(2009)}]{LambertLePoncin-Lafitte2009}
{Lambert} SB, {Le Poncin-Lafitte} C (2009) {Determining the relativistic
  parameter {$\gamma$} using very long baseline interferometry}. Astronomy \&
  Astrophysics 499:331--335
  
\bibitem[{{Lambert} and {Le
  Poncin-Lafitte}(2011)}]{LambertLePoncin-Lafitte2011}
{Lambert} SB, {Le Poncin-Lafitte} C (2011) {Improved determination of
  {$\gamma$} by VLBI}. Astronomy \& Astrophysics 529:A70
  
\bibitem[{Lebach et~al(1995)Lebach, Corey, Shapiro, Ratner, Webber, Rogers,
  Davis, and Herring}]{LebachCoreyShapiro1995}
Lebach DE, Corey BE, Shapiro II, et al. (1995) Measurement of the solar gravitational deflection of radio
  waves using very-long-baseline interferometry. Phys Rev Lett 75:1439--1442
  
\bibitem[{{Ma} {et~al.}(1986){Ma}, {Clark}, {Ryan}, {Herring}, {Shapiro},
  {Corey}, {Hinteregger}, {Rogers}, {Whitney}, {Knight}, {Lundqvist},
  {Shaffer}, {Vandenberg}, {Pigg}, {Schupler}, \& {Ronnang}}]{MaClarkRyan1986}
{Ma}, C., {Clark}, T.~A., {Ryan}, J.~W., {et~al.} 1986, Astronomical Journal, 92, 1020

\bibitem[{{Mignard} and {Klioner}(2009)}]{MignardKlioner2009}
{Mignard} F, {Klioner} S (2009) {{Gaia: Relativistic modelling and testing}}.
  In: {Klioner} S, {Seidelman} PK (eds) {{Relativity in Fundamental Astronomy
  -- Proceedings IAU Symposium No. 261}}, Cambridge University Press, vol~5, pp
  306--314

\bibitem[{Robertson and Carter(1984)}]{RobertsonCarter1984}
Robertson DS, Carter WE (1984) Relativistic deflection of radio signals in the
  solar gravitational field measured with vlbi. Nature 310:572
  
\bibitem[{Robertson et~al(1991)Robertson, Carter, and
  Dillinger}]{RobertsonCarterDillinger1991}
Robertson DS, Carter WE, Dillinger WH (1991) New measurement of solar
  gravitational deflection of radio signals using vlbi. Nature 349:768
  
\bibitem[{{Shapiro}(1964)}]{Shapiro1964}
{Shapiro} II (1964) {{Fourth Test of General Relativity}}. Phys Rev Lett
  13(26):789--791

\bibitem[{Shapiro(1967)}]{Shapiro1967}
Shapiro II (1967) New method for the detection of light deflection by solar
  gravity. Science 157(3790):806
  
 \bibitem[{{Shapiro} et~al(2004){Shapiro}, {Davis}, {Lebach}, and
  {Gregory}}]{ShapiroDavisLebach2004}
{Shapiro} SS, {Davis} JL, {Lebach} DE, {Gregory} JS (2004) {Measurement of the
  Solar Gravitational Deflection of Radio Waves using Geodetic
  Very-Long-Baseline Interferometry Data, 1979 1999}. Physical Review Letters
  92(12):121101

\bibitem[{{Titov}(2000)}]{2000ITN_Titov}
{Titov}, O. 2000, IERS Technical Note, 28, 11

\bibitem[{{Titov} {et~al.}(2004){Titov}, {Tesmer}, \&
  {Boehm}}]{2004ivsg.conf..267T}
{Titov}, O., {Tesmer}, V., \& {Boehm}, J. 2004, in International VLBI Service
  for Geodesy and Astrometry 2004 General Meeting Proceedings, ed. N.~R.
  {Vandenberg} \& K.~D. {Baver}, 267

\bibitem[{{Titov} and {Stanford}(2013)}]{TitovStanford2013}
{Titov} O, {Stanford} L (2013) {Comparison of wet troposphere variations
  estimated from VLBI and WVR}. In: {Zubko} N, {Poutanen} M (eds) 21st Meeting
  of the European VLBI Group for Geodesy and Astronomy, Reports of the Finnish
  Geodetic Institute, p. 151-154

\bibitem[{{Titov} et~al(2018){Titov}, {Girdiuk}, {Lambert}, {Lovell}, {McCallum}, {Shabala}, {McCallum}, {Mayer}, {Schartner}, {de Witt}, {Shu}, {Melnikov}, {Ivanov}, {Mikhailov}, {Yi}, {Soja}, {Xia}
and  {Jiang}}]{Titov2018}
{Titov} O, {Girdiuk} A, {Lambert} SB, et al. (2018) Testing general relativity with geodetic VLBI. What a single, specially designed experiment can teach us. Astronomy \& Astrophysics 618:A8

\bibitem[{{Titov} and {Lopez}(2018)}]{TitovLopez2018}
{Titov} O, {Lopez} Yu (2018) {Two-component structure of the radio source 0014+813 from VLBI observations within the CONT14 program}. Astronomy Letters, 44, 139


\bibitem[{{Will}(1993)}]{Will1993}
{Will} CM (1993) {Theory and Experiment in Gravitational Physics}

\end{thebibliography}

%---------------------------------------------------------------------------
%\begin{thebibliography}{99}
%--------------------------
%\providecommand{\natexlab}[1]{#1}
%\providecommand{\url}[1]{\texttt{#1}}
%\expandafter\ifx\csname urlstyle\endcsname\relax
%  \providecommand{\doi}[1]{doi: #1}\else
%  \providecommand{\doi}{doi: \begingroup \urlstyle{rm}\Url}\fi

%\bibitem[blabla et al.(2015)]{blabla_2015}
%Blabla~X, Bloblo~Z, Blublu~Y (2015)
%\newblock blabla blabla blabla blabla blabla blabla blabla.
%\newblock \emph{J~blabla}, 1, 1--99,
%\newblock \doi{blabla-blabla}.

%\bibitem[blublu et al.(2017)]{blublu_2017}
%Blublu~Y, Blabla~X, Bloblo~Z (2017)
%\newblock blublu blublu blublu blublu blublu blublu blublu blublu.
%\newblock \emph{J~blublu}, 1, 100--115,
%\newblock \doi{blublu-blublu}.

%\bibitem[bloblo et al.(2019)]{bloblo_2019}
%Bloblo~Z, Blublu~Y, Blabla~X (2019)
%\newblock bloblo bloblo bloblo bloblo bloblo bloblo bloblo.
%\newblock \emph{J~bloblo}, 1, 1--99,
%\newblock \doi{bloblo-bloblo}.

%-------------------
%\end{thebibliography}
%-------------------
%============================================================================
\end{document}